# Strain engineering of two-dimensional piezo-photocatalytic materials for hydrogen production

Zhao Liu[a], Biao Wang[a,b] and Claudio Cazorla[c*]

a. Research Institute of Interdisciplinary Science and School of Materials Science and Engineering, Dongguan University of Technology, Dongguan 523808, China
b. School of Physics and Sino-French Institute of Nuclear Engineering and Technology, Sun Yat-Sen University, Zhuhai 519082, China
c. Departament de Física, Universitat Politècnica de Catalunya, Campus Nord B4-B5, Barcelona 08034, Spain; Barcelona Research Center in Multiscale Science and Egineering, Universitat Politècnica de Catalunya, Barcelona 08019, Spain

* Corresponding author: claudio.cazorla@upc.edu

**Abstract**

Low-dimensional transition metal dichalcogenides (TMDC) exhibit great photocatalytic performance and tunability. In this work, using first-principles simulations based on density functional theory (DFT), we demonstrate that external electric bias can be employed to further improve the photocatalytic hydrogen production efficiency of the six $AB_2$ (A=Mo, W and B=S, Se, Te) TMDC monolayers by exploiting their piezoelectric response. In particular, when subjected to a proper amount of electrically induced tensile biaxial strain, most TMDC monolayers turn into potentially ideal photocatalyst towards the hydrogen evolution reaction (HER). The beneficial effects of introducing tensile biaxial strain on the TMDC monolayers are not limited to the reduction of the band gap and proper adjustment of the band edge positions, but also to the modification of the H adsorption free energy in such a way that the HER reaction is noticeably favored.

**Introduction**

At present, due to the growing environmental problems and energy crisis, seeking alternative energy resources to fossil fuels has become critically important [1-3]. Hydrogen production through water splitting by photocatalytic materials is one of the most promising techniques because converting solar energy into chemical energy by photocatalysis is clean, sustainable and cost-effective [4-6]. Ever since Fujishima *et al.*'s discovery of the splitting of water into hydrogen and oxygen by UV light-induced electrocatalysis using a titanium dioxide photoanode [7], the photocatalytic performance of various kinds of bulk materials have been investigated, including $TiO_2$, $CeO_2$, ZnO and $WO_3$ [8-16]. Despite many efforts, however, most three-dimensional bulk photocatalysts exhibit low photocatalytic efficiency. In contrast, two-dimensional materials have larger surface areas and shorter electron/hole transport distances as compared with their bulk counterparts, making them good candidates as photocatalysts for hydrogen production by water splitting [17-19].

Among the many kinds of 2D materials, transition metal dichalcogenides (TMDCs) have been extensively studied for their unique chemical and physical properties, which enable them to be applied in different fields, such as nanoelectronics, photonics, opto-electronics, sensing, energy storage and hydrogen production [20-24]. However, for the particular aim of photocatalysis, TMDCs are not optimal from an applied point of view because their band gaps and band edge positions do not fulfill the requirements of an ideal photocatalyst [25-27]. In particular, the band gap of an ideal photocatalyst should be within the range of 1.23-3.00 eV, so that the absorption of visible light is efficient, and the corresponding valence band maximum (VBM) should be lower in energy than the hydrogen evolution reaction (HER) potential while simultaneously the conduction band minimum (CBM) should be higher than the oxygen evolution reaction (OER) potential [28, 29].

Fortunately, TMDCs are well known for their great tunability and various techniques have been adopted to improve their optoelectronic properties [30], among which strain engineering has been demonstrated to be one of the most effective strategies both experimentally and theoretically [31-33]. Nevertheless, strain engineering involves complex synthesis techniques, like the growing of thin films on top of substrates and/or mechanical nanoactuators methods, that may hamper the development of novel photocatalytic applications based on TMDCs [34-36].

A possibility to take full advantage of strain engineering techniques for TMDC-driven hydrogen sunlight production, by getting rid of such intricate synthesis/mechanical actuation methods, may consist in exploiting external electric fields [37-39]. Most TMDC single layers are piezoelectric materials (i.e., their crystal structure does not fulfill inversion crystal symmetry – are non-centrosymmetric) and consequently react to external electric field bias by elongating/contracting their lattice parameters. In this work, we explore this possibility by studying the photocatalytic and piezoelectric properties of six different $AB_2$ (A=Mo, W and B=S, Se, Te) TMDC monolayers under broad epitaxial strain conditions, which can be achieved in practice by applying external electric fields on them.

**Computational Methods**

All the first-principles calculations based on density functional theory were conducted using VASP [40]. Projected augmented plane-wave method was used to represent the ionic cores with the following electrons been considered as valence: Mo ($4p$, $5s$, $4d$), W ($5p$, $5s$, $5d$), S ($3s$, $3p$), Se ($4s$, $4p$) and Te ($5s$, $5p$) [41]. PBE functional within the GGA framework was used to describe the exchange-correlational interactions [42] while HSE06 hybrid functional was used for precise calculation of the electronic properties [43]. The plane wave cutoff energy of basis set was set to be 600 eV for all the analyzed compounds. Structural relaxations were performed with a quasi-Newton algorithm [44] in which the convergence threshold for the electronic energy and atomic forces were of $1 \times 10^{-7}$ eV and $1 \times 10^{-3}$ eV·Å$^{-1}$, respectively. Monkhorst-Pack k-point grids of spacing $2\pi \times 0.01$ Å$^{-1}$ were used for integration within the first Brillouin zone [45]. For a better description of the localized $d$ orbitals, a Hubbard-type on-site correction energy was applied on the Mo and W atoms with effective values of $U_{eff}$ 4.0 and 6.2 eV, respectively [46].

Periodic boundary conditions were applied along all the lattice vector directions and a vacuum region of 15 Å thickness was set along the out-of-plane direction to minimize the interactions between periodic single-layer images [47]. Biaxial strain was defined as $\eta = (a-a_0)/a_0$ [13], where $a_0$ is the in-plane lattice parameter of the fully relaxed structures. 3 × 3 supercells were used for the estimation of H adsorption free energies; the D3 Grimme correction method was adopted in these cases to address the long-range Van der Waals interactions [48]. For accurate estimation of the band alignments, we calculated the electrostatic potential of the VBM in the TMDC monolayers and the corresponding vacuum level. The absolute energy level of the VBM relative to vacuum was defined as the difference between the VB energy and the vacuum level; meanwhile, the CBM energy was determined by adding the band gap to the previously calculated VBM energy (at the HSE06 level).

In addition to the band gap and band alignments, we also evaluated the energetics of the hydrogen evolution reaction (HER) as mediated by TMDCs monolayers [49]. To do this, we calculated the adsorption free energy of one hydrogen atom ($\Delta G_H$) on the monolayer surfaces, which was estimated like [49]:

$$\Delta G_H = \Delta E_H + \Delta E_{ZPE} - T\Delta S \qquad (1)$$

where $\Delta E_H$ is the hydrogen adsorption energy, $\Delta E_{ZPE}$ and $\Delta S$ are the difference of zero-point energy and entropy between the adsorbed atomic hydrogen and hydrogen in the gas phase molecule, respectively. The contribution from the catalysts to both $\Delta E_{ZPE}$ and $\Delta S$ are very small and thus were neglected. The $\Delta E_{ZPE}$ term was calculated as the difference between the zero-point energy of the adsorbed hydrogen atom and one half of the zero-point energy of a $H_2$ gas molecule, namely

$$\Delta E_{ZPE} = E_{ZPE}^{H} - \frac{1}{2} E_{ZPE}^{H_2} \qquad (2)$$

It is worth noting that the value of the $\Delta E_{ZPE}$ term usually is quite small, of the order of 0.01 eV [50]. Meanwhile, $\Delta S$ was taken equal to minus one half of the experimental entropy of a $H_2$ gas molecule under standard thermodynamic conditions (i.e., 298K and 1 atm), which for room-temperature leads to $T\Delta S = -0.2$ eV [51].

**Results and Discussion**

Figure 1a shows the atomic structure of the analyzed $AB_2$ monolayer systems. We investigated the 1T polymorph for $MoS_2$ and the 2H polymorph for the other five compounds. $MoS_2$ naturally is thermodynamically stable in the 2H phase [52]; however, exposure of catalytically active edge Mo ions is limited in this phase [53, 54] and the corresponding electrical conductivity is also very low, thus resulting in unsatisfactory HER performance [55]. By contrast, 1T $MoS_2$ has experimentally demonstrated outstanding catalytic performance and several facile chemical routes have been recently proposed to synthesize this polymorph [56]. For these reasons, we selected 1T $MoS_2$ in the present study while considered the typically synthesized 2H polymorph for the rest of monolayers.

Figure 1a shows the atomic configuration of 1T-$MoS_2$ from the side and top views, respectively. This polymorph exhibits trigonal crystal symmetry (P−3m1 space group) with the Mo atoms intercalated between two S layers; our calculated lattice constant amounts to 3.19 Å. Figure 1a also shows the atomic structure of the 2H polymorph considered for the other five $AB_2$ monolayers. This phase has hexagonal crystal symmetry (P-6m2) and the A transition-metal ions also are intercalated between two B chalcogenide layers. The calculated lattice constants of $MoSe_2$, $MoTe_2$, $WS_2$, $WSe_2$ and $WTe_2$ are 3.35, 3.53, 3.18, 3.31 and 3.54 Å, respectively, which are in good agreement with those reported in previous works [57].

Both the 1T and 2H phases are non-centrosymmetric and consequently are piezoelectric. Thus, by applying external electric fields in them in principle is possible to achieve dynamic control, instead of static, of the in-plane biaxial strain thanks to their piezoelectric response. Figure 1b shows the calculated lattice strain in 1T $MoS_2$ expressed as a function of applied electric field along the lattice *b* direction. It is shown that both lattice parameters *a* and *b* experience a drastic change when a small electric field of ~0.1 eV/Å is applied, *a* being increased and *b* shortened (the sign of these variations are related to the sign of the piezoelectric coefficient $\varepsilon_{yy}$, see below). For lager electric field values, the rate of variation of the *a* and *b* lattice parameters are practically linear and of the same sign. The estimated lattice parameter change can be as high as 1% for an electric field of ~ 1 eV/Å, thus application of electric bias can be an efficient means to introduce structural strains in TMDC monolayers with the aim of tuning their physicochemical properties. Explicit simulation of electric fields in first-principles calculations, however, is computationally expensive and intricate [58, 59]. Therefore, for the next photocatalytic analysis to be presented, we straightforwardly considered biaxial lattice strains of arbitrary size in our simulation cells while disregarded the presence of the originating electric fields.

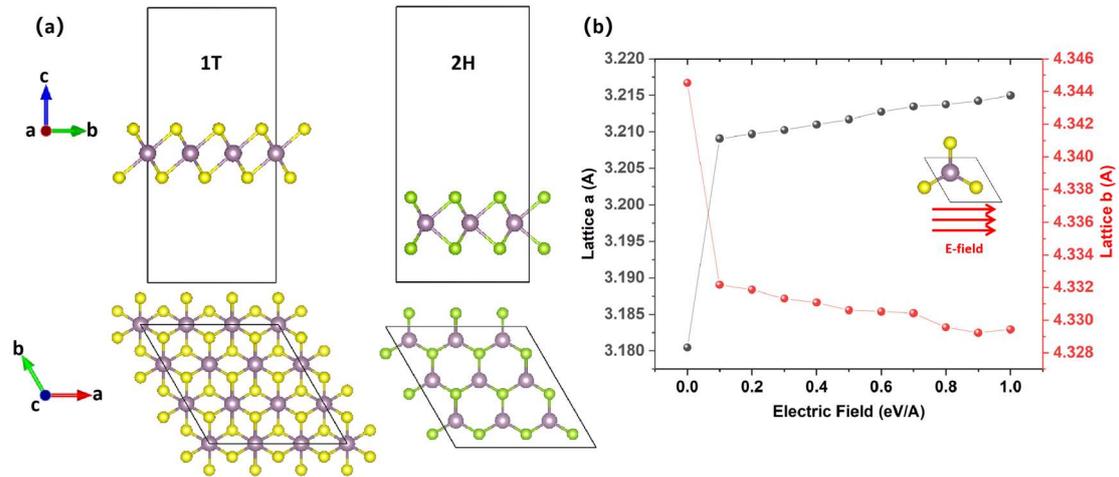

**Figure 1.** (a) The atomic structure of the investigated TMDC monolayer systems. (b) The estimated structural strain induced in 1T $MoS_2$ by means of external electric fields. The results were obtained with the semi-local PBE DFT functional.

Table 1 lists the calculated piezoelectric coefficient $\varepsilon_{yy}$ expressed as a function of epitaxial strain for the six considered TMDC monolayers. In the absence of strain, the reported piezoelectric constants in general are negative (i.e., the *b* lattice parameter will shrink under the action of an electric field applied along the *y* direction) and small in absolute value. However, interestingly, under increasing epitaxial strain the absolute values of the piezoelectric constants increase, thus under the action of an electric bias the piezoelectric response of TMDC monolayers is enhanced; in some cases, the sign of $\varepsilon_{yy}$ also changes. For instance, at zero strain WSe$_2$ exhibits a positive piezoelectric coefficient while the other five monolayers present negative $\varepsilon_{yy}$ values; however, under a compressive strain of 7% the piezoelectric coefficient of 1T-MoS is the only one that is positive and under a tensile strain of 7% strain MoS$_2$, WSe$_2$ and WTe$_2$ exhibit positive $\varepsilon_{yy}$ values. In addition to the change in sign of $\varepsilon_{yy}$, the absolute values of all piezoelectric coefficients significantly increase as referred to the zero-strain value either under compressive or tensile strains (e.g., by one order of magnitude in MoS$_2$, WS$_2$ and WTe$_2$). These results suggest that biaxial strains of about 1% could be realized in TMDC monolayers by applying relatively small electric bias on them.

**Table 1.** Calculated in-plane piezoelectric coefficient $\varepsilon_{yy}$ (C/m$^2$) in TMDC monolayers expressed as a function of epitaxial strain. The results were obtained with the semi-local PBE DFT functional.

| Materials | -7% | 0 | 7% |
| --- | --- | --- | --- |
| MoS$_2$ (1T) | 0.00193 | -0.00022 | 0.00219 |
| MoSe$_2$ (2H) | -0.25576 | -0.11762 | -0.32265 |
| MoTe$_2$ (2H) | -0.2033 | -0.14236 | -0.08272 |
| WS$_2$ (2H) | -0.66089 | -0.0738 | -0.08104 |
| WSe$_2$ (2H) | -0.02478 | 0.01314 | 0.05292 |
| WTe$_2$ (2H) | -0.09828 | -0.01152 | 0.01296 |

Figure 2 shows the electronic band gap of the analyzed TMDC monolayers expressed as a function of epitaxial strain. 1T-MoS$_2$ turns out to be metallic under all the investigated strains, thus is not explicitly shown. Under zero strain, the band gaps of all five compounds are approximately in the range of 1.6–2.1 eV, with WTe$_2$ presenting the smallest (1.57 eV) and WSe$_2$ the largest (2.09 eV). These band gaps fulfill the requirement of ideal photocatalysts (see above). However, further reduction of the band gap could potentially improve the TMDC monolayers photocatalytic efficiency thanks to enhanced light absorption. According to our calculations, tensile biaxial strain can effectively reduce the band gap energy of all five compounds. The band gap reduction in the WS$_2$ monolayer can be as high as ~72% when a +7% strain is applied on it. The other four compounds also show a band gap reduction of more than 40% for the largest tensile biaxial strain considered in this study. In contrast, compressive biaxial strain does not induce a systematic band gap change in the five compounds. For instance, under a -7% biaxial strain MoSe$_2$ and WS$_2$ present a band gap increase of ~25% with respect to their zero-strain values while MoTe$_2$ and WTe$_2$ exhibit band gap reductions that are similar in size to those observed under tensile strain.

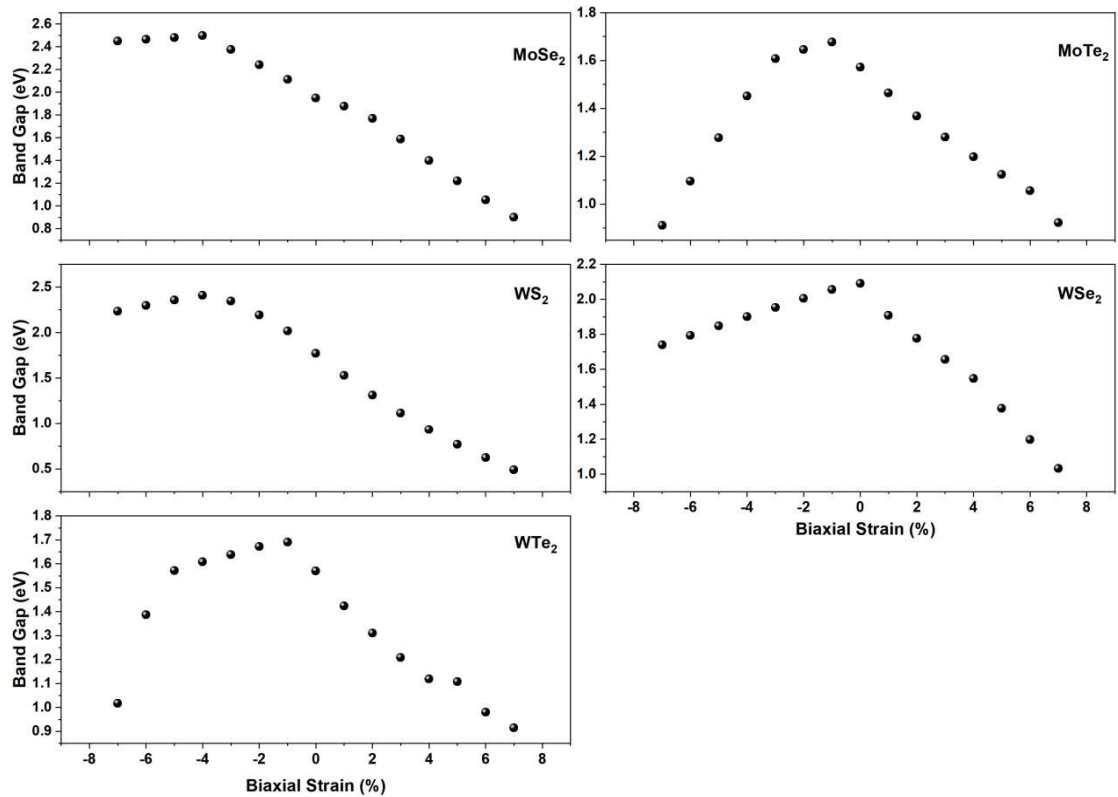

**Figure 2.** Calculated energy band gap evolution of TMDC monolayers expressed as a function of epitaxial strain. The results were obtained with the hybrid HSE06 DFT functional.

In order to understand the physical causes underlying the strain-induced band gap variations shown in Figure 2 and to further analyze the influence of biaxial strain on the optoelectrical properties of TMDC monolayers, we also studied their band structure in detail. Figure 3 shows the results of our band structure calculations expressed as a function of epitaxial strain (cases of -7%, 0 and 7%). 1T-$MoS_2$ shows metallic behavior in all the considered cases, thus does not seem to be a suitable candidate for photocatalytic hydrogen production applications due to the lack of a band gap and overpotential for water splitting redox potential. Nevertheless, due to its metallic nature, 1T-$MoS_2$ still can be a good electrocatalytic material for water splitting (see below). The other five monolayers all exhibit direct semiconducting behavior in the absence of strain. However, under either compressive or tensile strain, all five monolayers exhibit a direct to indirect band-gap transition. This electronic band-gap transformation can be attributed to the symmetry breaking triggered by the biaxial strain. Our crystal symmetry analysis in fact confirms that the when the five monolayers are biaxially strained they all experience a symmetry breaking from hexagonal (space group 187) to orthorhombic (space group 38). As a consequence, the VBM and CBM levels shift away from the K point, leading to the observed band-gap transformation.

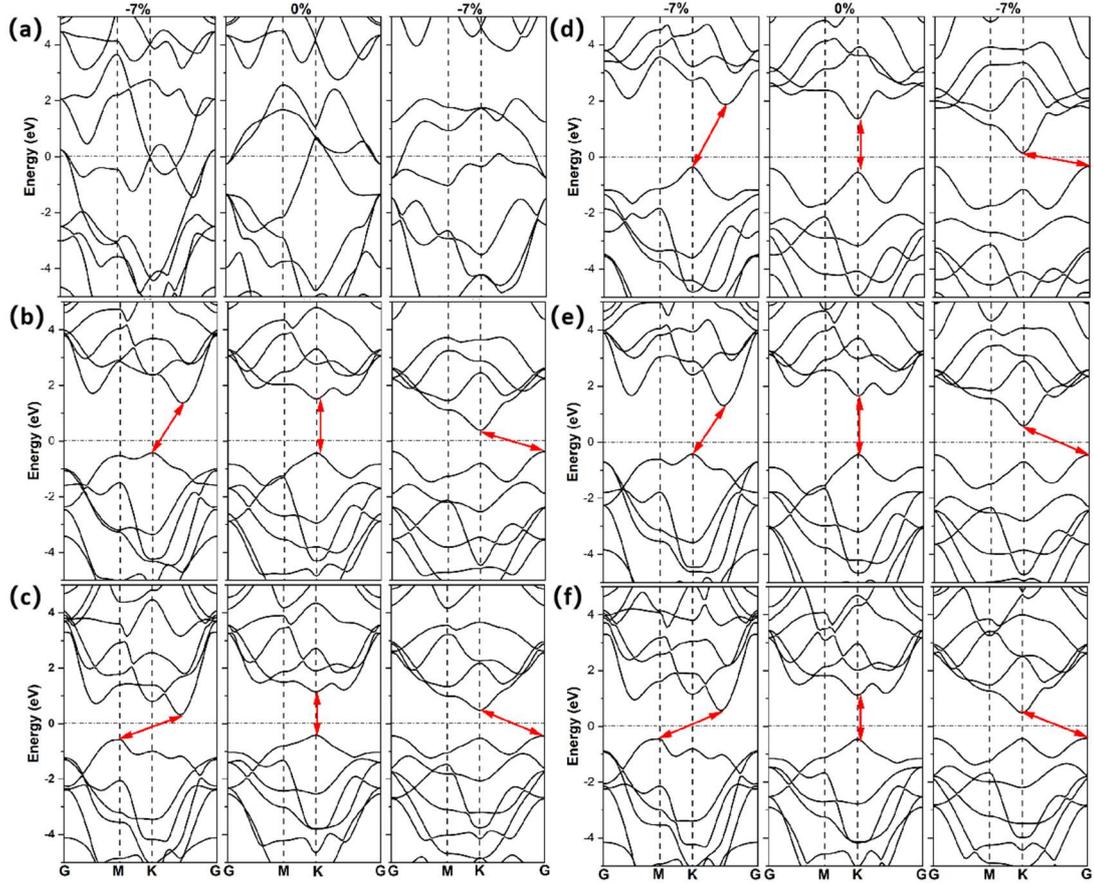

**Figure 3.** Calculated band structure of TMDC monolayers calculated as a function of epitaxial strain. (a) $MoS_2$, (b) $MoSe_2$, (c) $MoTe_2$, (d) $WS_2$, (e) $WSe_2$, (f) $WTe_2$. The results were obtained with the hybrid HSE06 DFT functional.

To clearly visualize the direct to indirect band-gap transformation, we took $MoSe_2$ as a case to study its 3D band structure. Figure 4 shows the 3D band structure of $MoSe_2$ obtained under three different strain conditions. In all the cases, the VBM and CBM correspond to the $13^{th}$ and $14^{th}$ electronic bands, respectively. In the absence of biaxial strain, the top of the valence band and bottom of the conduction band are located at the same *k*-point, K, resulting in a direct band gap. Under 7% of compressive strain, the CBM shifts away to a non-high symmetric point, while the VBM remains at the M point, thus leading to an indirect band gap. Similarly, under 7% of tensile strain, the VBM shifts to the *G* point while the CBM remains at the *K* point, which also leads to an indirect band gap.

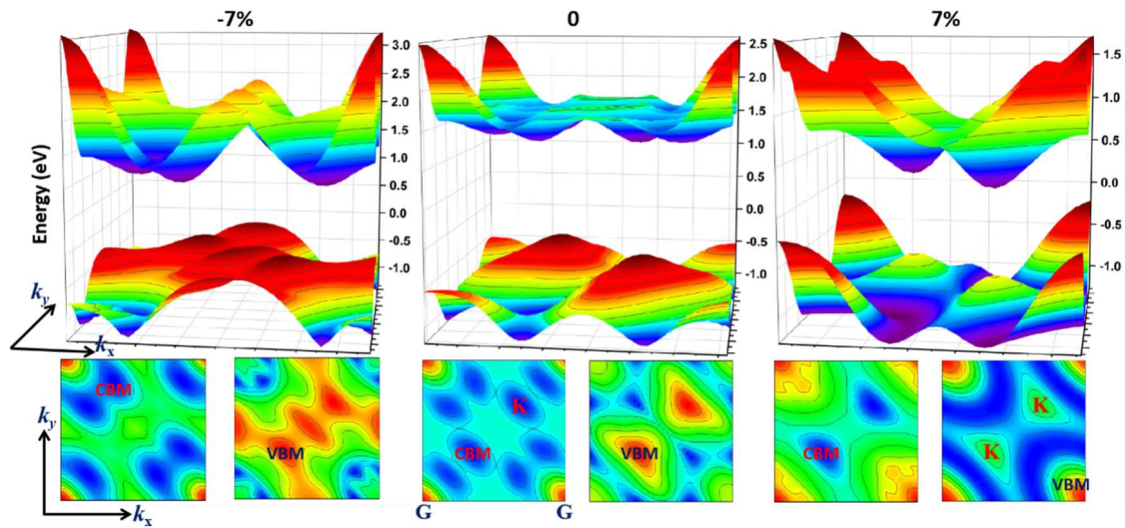

**Figure 4.** The 3d band structure of the MoSe2 at different strains. The bottom panel of each figure shows the 2d contour map of the 13$^{th}$ and 14$^{th}$ band, where the position of the CBM and VBM are labelled correspondingly. The results were obtained with the semi-local PBE DFT functional.

Figure 5 shows the calculated band alignments of the five TMDC monolayers expressed as a function of epitaxial strain. Let us start commenting the special case of $WS_2$. When no strain is applied, the band alignments of this monolayer are located at -6.1 eV and -4.34 eV, respectively, which indicates that unstrained $WS_2$ is already suitable for driving the water splitting reaction and generate molecular hydrogen. Under compressive biaxial strain, the band edge positions of $WS_2$ are slightly elevated, making the VBM closer to the oxidation potential; however, compressive strain also increases the $WS_2$ band gap (e.g., 2.23 eV at -7% strain), which is not good for light absorption purposes. In contrast, tensile strain, which also induces a slight increase of the VBM level, significantly reduces the band gap of $WS_2$ by lowering its CBM level, which suggests that the photocatalytic activity of $WS_2$ could be optimized in practice by introducing a small amount of tensile biaxial strain on it.

Besides $WS_2$, the VBMs of the other four TMDC monolayers are all situated higher than the oxidation potential of oxygen, which makes them unsuitable for photocatalytic water splitting due to insufficient over potential for the oxygen oxidation reaction. Interestingly, under increasing tensile biaxial strain, the VBM levels of MoTe$_2$ and WTe$_2$ are systematically reduced and those of MoSe$_2$ and WSe$_2$ reach their minimum at +2% and +4%, respectively. The CBM levels of all five monolayers generally decrease under tensile biaxial strain, thus resulting in an overall reduction of the band gap that is consistent with the results shown in Figure 2. Thus, in all the analyzed TMDC monolayers, the effects of tensile strain on the band alignments are overall beneficial, although in different measures.

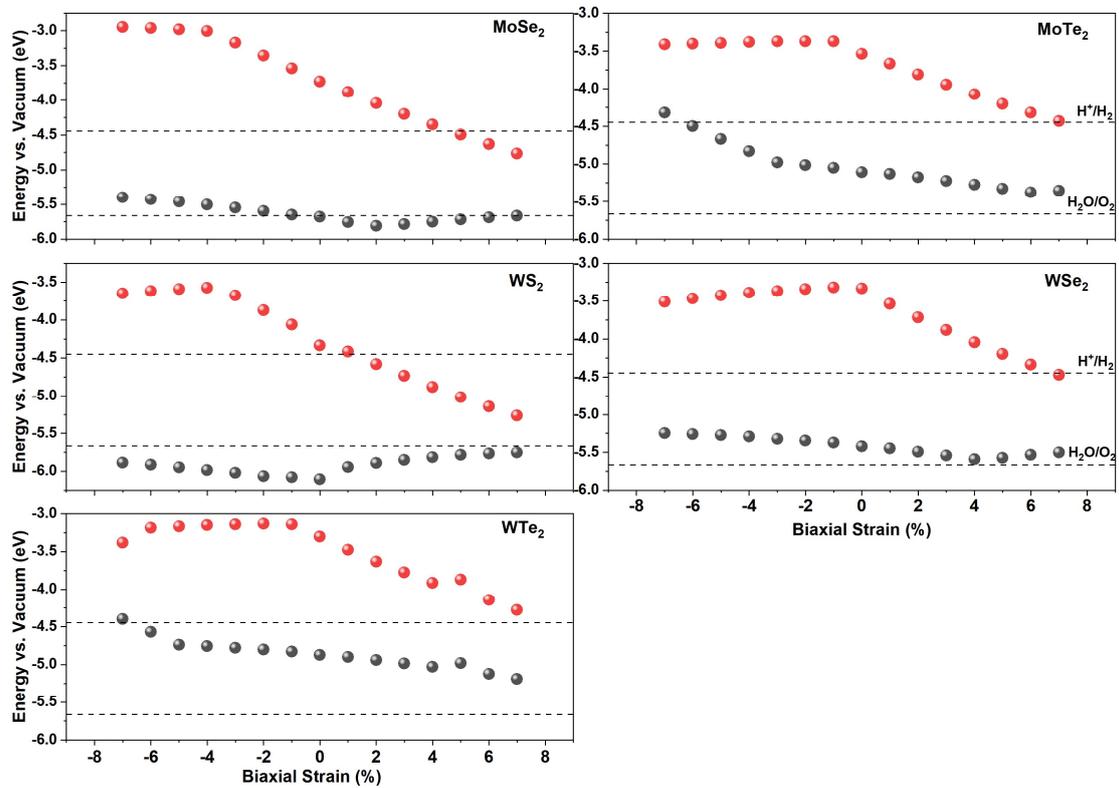

**Figure 5.** Calculated band alignments of TMDC monolayers estimated as a function of epitaxial strain. Red and black dots represent the CBM and VBM levels, respectively. The results were obtained with the hybrid HSE06 DFT functional.

Figure 6 shows the H adsorption free energies, $\Delta G_H$, calculated for the 6 TMDC monolayers and expressed as a function of epitaxial strain. At zero strain, the hydrogen adsorption energies are all positive and large (i.e., of the order of 1-2 eV), thus suggesting that the interactions between H atoms and the TMDC monolayers are repulsive and moderately strong. 1T-MoS$_2$ presents the smallest $\Delta G_H$ value hence despite of its unsuitability to drive the full water splitting reaction due to its metallic character it can be the most appropriate for HER purposes. Meanwhile, it is expected that any strategy that can reduce the sizeable $\Delta G_H$ values estimated for unstrained TMDC monolayers will improve their HER performances. Figure 6 clearly shows that tensile biaxial strain can be used to consistently decrease the value of the H adsorption energies, thus potentially improving the HER performance of all TMDC monolayers. For instance, under a tensile strain of +7% the achieved reduction in $\Delta G_H$ as referred to the equilibrium state turns out to be of ≥30% in all the analyzed cases.

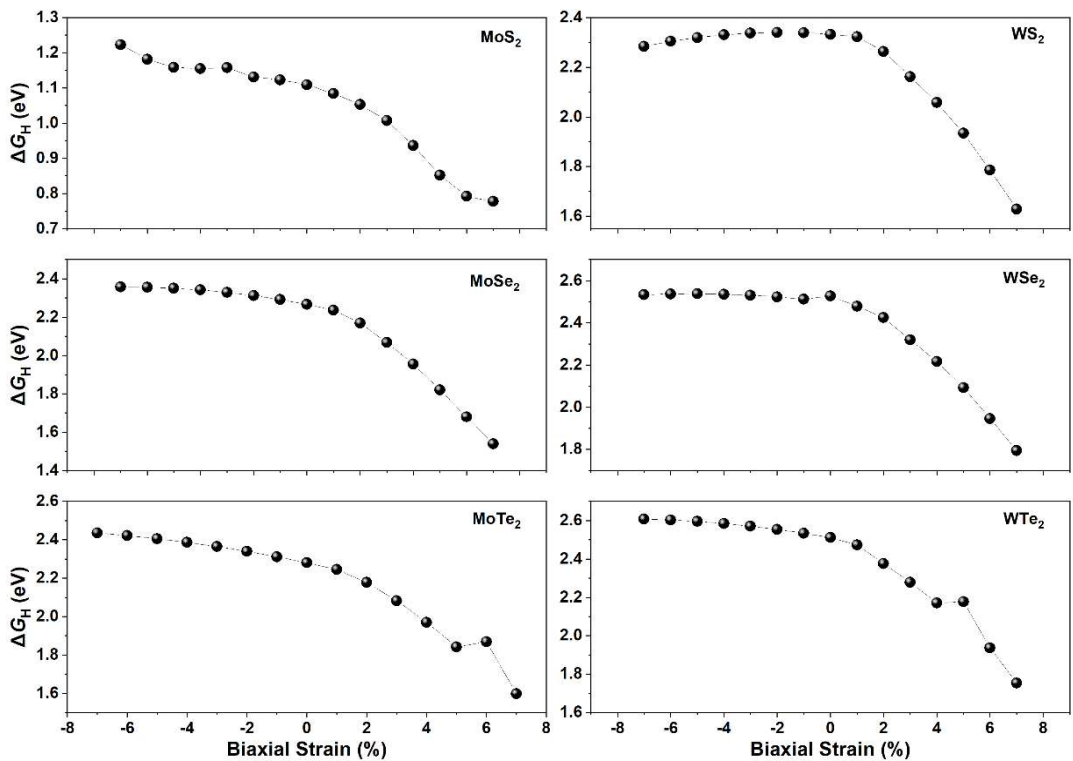

**Figure 6.** Calculated H adsorption free energies expressed as a function of epitaxial strain for the six considered monolayers. The results were obtained with the semi-local PBE DFT functional.

To further unveil the influence of biaxial strain on the H adsorption free energies of TMDC monolayers, we analyzed the MoSe$_2$ system in detail since this is a representative case. Figure 7a shows the charge density difference calculated for the H atom adsorbed on the TMDC monolayer, taking the free H atom as the reference system. The cyan isosurface indicates charge depletion, namely, the H atom transfers electronic charge to the Se atom to which is directly bonded to (see yellow isosurface therein). This effect is due to the higher electronegativity of Se as compared to that of H and occurs at any strain. Nevertheless, in this representation the differences in H charge depletion between the three analyzed strain cases are quite small and thus it is difficult to infer any clear dependence of this effect on biaxial strain. To better understand these results, we conducted complementary charge analysis of the adsorbed H atom under different strain conditions, the results of which are shown in Figure 7b. In particular, we numerically estimated the charge difference in the H atom considering the TMDC and free atom cases, $\Delta Q_H$. Negative (positive) values of this quantity indicate electronic charge depletion (accumulation) as compared to the free atom case.

At zero strain, the charge difference in the H atom adsorbed on the MoSe$_2$ monolayer is negative and approximately equal to 0.14e in absolute value, thus indicating charge depletion as compared to the free atom case (in consistent agreement with the charge density difference results shown in Figure 7a). When charge is transferred from the H atom to the Se atom that is underneath of it, the first gets positively charged and the second negatively; as a consequence, a surface dipole is created and attractive interactions of Coulomb type act between the two ions. Thus, the larger the amount of charge that is transferred between the H and Se atoms in principle the more attractive the electrostatic interactions between them should be and the smaller the value of $\Delta G_H$.

Under compressive strain, the absolute value of $\Delta Q_H$ slightly decreases and the minus sign is conserved, hence in comparison to the unstrained case a bit less of charge is transferred from the H to the chalcogenide atom. Consequently, the electrostatic H-Se attraction diminishes and $\Delta G_H$ somewhat increases, as it is shown in Figure 6. At the maximum compressive strain considered in this work, $\Delta Q_H$ is practically identical in size to that calculated under no strain hence the H adsorption modes should be very similar in the two cases. This latter point is corroborated by the H-Se distances shown in Figure 7a since the one corresponding to the -7% strain case (1.55 Å) is practically the same as that obtained under equilibrium conditions (1.56 Å).

Conversely, the steady $\Delta G_H$ decrease observed under increasing tensile strain (Figure 6) can be understood in terms of enhanced electrostatic interactions between the H atom and MoSe$_2$ surface. As it is shown in Figure 7b, under increasing tensile strain the amount of charge donated by the H atom rapidly increases as conveyed by the almost linear growth in absolute value of the quantity $\Delta Q_H$. For instance, in the +6% biaxial strain case the relative H charge change is roughly equal to -0.16e; likewise, the H-Se distance now is reduced to 1.49 Å. Reassuringly, Figure 7c shows explicitly the strong correlation between the two quantities $\Delta G_H$ and $\Delta Q_H$, which as explained here can be rationalized in simple terms of practically constant and enhanced electrostatic Coulomb interactions under compressive and tensile biaxial strains, respectively.

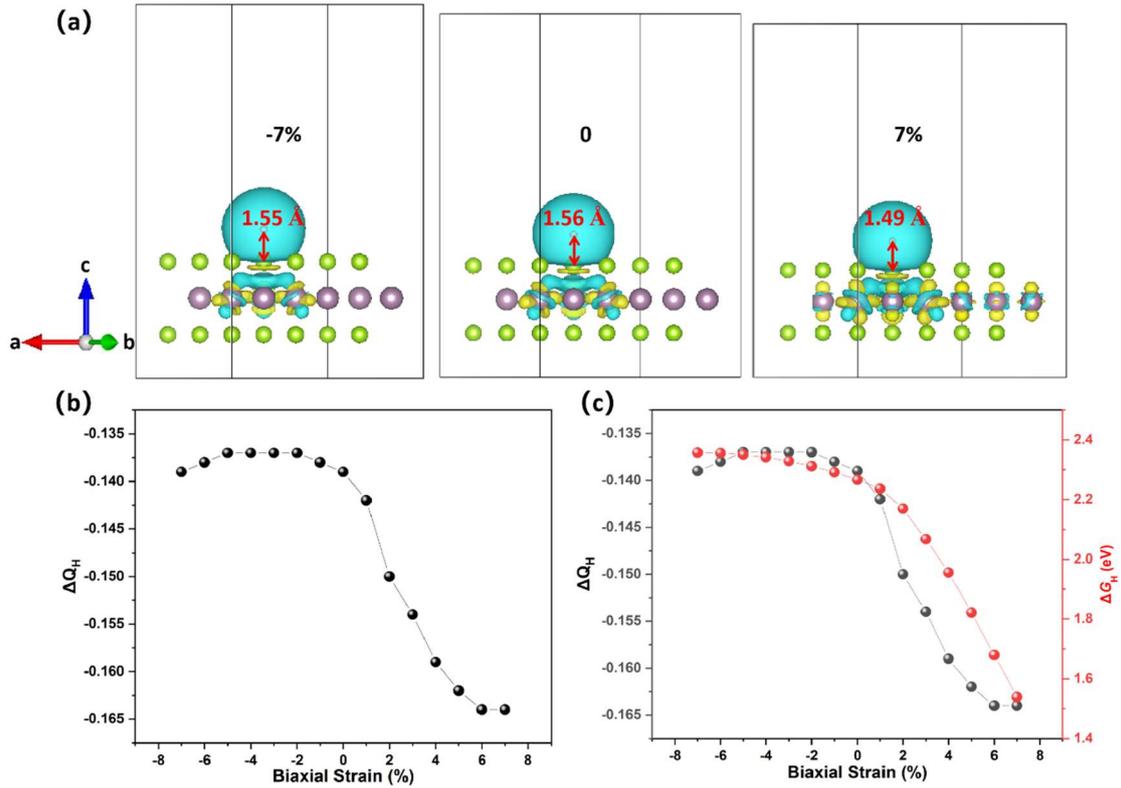

**Figure 7.** (a) Charge density difference of the MoSe$_2$ monolayer with an adsorbed H atom under different biaxial strains. The cyan lobe represents charge depletion while the yellow one represents charge accumulation; the isosurface for the charge density difference is 0.003 electrons/Bohr$^3$ for all the cases; (b) Charge analysis of the adsorbed H atom as a function of epitaxial strain in MoSe$_2$. ΔQ$_H$ is defined as the electronic charge difference between the H atom on the TMDC surface and the free H atom. (c) Comparison of the ΔQ$_H$ and Δ$G_H$ as a function of epitaxial strain.

**Conclusions**

We have demonstrated by means of first-principles simulation methods that electric field driven biaxial strain can be used to improve the HER performance of TMDC monolayers by tuning their optoelectronic and photocatalytic properties in a systematic and predictable manner. The piezoelectric constants of the monolayers increase under biaxial strain, thus suggesting that in principle it is possible to introduce considerable biaxial strains in the monolayers by using small electric fields. 1T-MoS$_2$ shows metallic behavior under all considered strain cases hence it is not suitable for driving photocatalytic hydrogen production out of water due to the lack of over potential. In contrast, the zero-strain band gaps of the other five monolayers vary from 1.5 eV to 2 eV and tensile biaxial strain can be used to reduce them systematically. In addition, our band alignment calculations suggest that at equilibrium conditions most TMDC monolayers are not suitable for photocatalytic water splitting since their VBM levels are higher than the oxidation potential for oxygen. However, tensile biaxial strain again can lower those VBM levels in a systematic and beneficious manner. Furthermore, the HER performance of all analyzed TMDC monolayers can be significantly enhanced also by means of tensile strain. Therefore, by introducing a proper amount of biaxial strain on the TMDC monolayers they can be potentially turned into ideal photocatalysts with suitable band gaps, proper band edge positions and increased HER activity. Our work demonstrates that strain engineering of TMDC piezoelectric monolayers, facilitated by the application of external electric fields, represents an efficient and very promising strategy for developing novel water splitting photocatalytic materials. This work may drive and motivate extensive experimental efforts for better rational design of TMDC photocatalysts for hydrogen production.

**Declaration of competing interest**

The authors declare no competing financial interests.

**Acknowledgments**

This work was supported by the National Natural Science Foundation of China (12002402, 11832019) and the NSFC original exploration project (12150001). The authors also appreciate the financial support from the Project of Nuclear Power Technology Innovation Center of Science Technology and Industry for National Defense (HDLCXZX-2019-ZH-31). This work was also supported by the Guangdong

International Science and Technology Cooperation Program (2020A0505020005). Z L wants to thank the financial support from the Guangdong overseas young postdoctors recruitment program. C.C. acknowledges support from the Spanish Ministry of Science, Innovation and Universities under the "Ramón y Cajal" fellowship RYC2018-024947-I.

**References**

1. Eisa, M., et al., *Role and Responsibility of Sustainable Chemistry and Engineering in Providing Safe and Sufficient Nitrogen Fertilizer Supply at Turbulent Times.* 2022, ACS Publications. p. 8997-9001.
2. Xie, W., et al., *Toward the next generation of sustainable membranes from green chemistry principles.* ACS Sustainable Chemistry & Engineering, 2020. **9**(1): p. 50-75.
3. Myung, Y., et al., *Graphene-based aerogels derived from biomass for energy storage and environmental remediation.* Acs Sustainable Chemistry & Engineering, 2019. **7**(4): p. 3772-3782.
4. Khaselev, O. and J.A. Turner, *A monolithic photovoltaic-photoelectrochemical device for hydrogen production via water splitting.* Science, 1998. **280**(5362): p. 425-427.
5. Hisatomi, T. and K. Domen, *Reaction systems for solar hydrogen production via water splitting with particulate semiconductor photocatalysts.* Nature Catalysis, 2019. **2**(5): p. 387-399.
6. Nann, T., et al., *Water splitting by visible light: a nanophotocathode for hydrogen production.* Angewandte Chemie International Edition, 2010. **49**(9): p. 1574-1577.
7. Fujishima, A. and K. Honda, *Electrochemical photolysis of water at a semiconductor electrode.* nature, 1972. **238**(5358): p. 37-38.
8. Nakata, K. and A. Fujishima, *TiO2 photocatalysis: Design and applications.* Journal of photochemistry and photobiology C: Photochemistry Reviews, 2012. **13**(3): p. 169-189.
9. Guo, Q., et al., *Fundamentals of TiO2 photocatalysis: concepts, mechanisms, and challenges.* Advanced Materials, 2019. **31**(50): p. 1901997.
10. Liu, Z., et al., *Planar-dependent oxygen vacancy concentrations in photocatalytic CeO 2 nanoparticles.* CrystEngComm, 2018. **20**(2): p. 204-212.
11. Liu, Z., et al., *Growth mechanism of ceria nanorods by precipitation at room temperature and morphology-dependent photocatalytic performance.* CrystEngComm, 2017. **19**(32): p. 4766-4776.
12. Liu, Z., et al., *DFT study of methanol adsorption on defect-free CeO2 low-index surfaces.* ChemPhysChem, 2019.
13. Liu, Z., et al., *Strain engineering of oxide thin films for photocatalytic applications.* Nano Energy, 2020: p. 104732.
14. Qi, K., et al., *Review on the improvement of the photocatalytic and antibacterial activities of ZnO.* Journal of Alloys and Compounds, 2017. **727**: p. 792-820.

15. Yu, D., et al., *Solar Photocatalytic Oxidation of Methane to Methanol with Water over RuOx/ZnO/CeO2 Nanorods.* ACS Sustainable Chemistry & Engineering, 2021. **10**(1): p. 16-22.
16. Tang, H., et al., *Visible-light localized surface plasmon resonance of WO3–x nanosheets and its photocatalysis driven by plasmonic hot carriers.* ACS Sustainable Chemistry & Engineering, 2021. **9**(4): p. 1500-1506.
17. Luo, B., G. Liu, and L. Wang, *Recent advances in 2D materials for photocatalysis.* Nanoscale, 2016. **8**(13): p. 6904-6920.
18. Su, Q., et al., *Heterojunction photocatalysts based on 2D materials: the role of configuration.* Advanced Sustainable Systems, 2020. **4**(9): p. 2000130.
19. Singh, A.K., et al., *Computational screening of 2D materials for photocatalysis.* The journal of physical chemistry letters, 2015. **6**(6): p. 1087-1098.
20. Gupta, U. and C. Rao, *Hydrogen generation by water splitting using MoS2 and other transition metal dichalcogenides.* Nano Energy, 2017. **41**: p. 49-65.
21. Mak, K.F. and J. Shan, *Photonics and optoelectronics of 2D semiconductor transition metal dichalcogenides.* Nature Photonics, 2016. **10**(4): p. 216-226.
22. Wang, Q.H., et al., *Electronics and optoelectronics of two-dimensional transition metal dichalcogenides.* Nature nanotechnology, 2012. **7**(11): p. 699-712.
23. Wang, L., et al., *Transition metal dichalcogenides for sensing and oncotherapy: status, challenges, and perspective.* Advanced Functional Materials, 2021. **31**(5): p. 2004408.
24. Yu, P., et al., *Earth abundant materials beyond transition metal dichalcogenides: A focus on electrocatalyzing hydrogen evolution reaction.* Nano Energy, 2019. **58**: p. 244-276.
25. Gong, C., et al., *Band alignment of two-dimensional transition metal dichalcogenides: Application in tunnel field effect transistors.* Applied Physics Letters, 2013. **103**(5): p. 053513.
26. Chiu, M.H., et al., *Band alignment of 2D transition metal dichalcogenide heterojunctions.* Advanced Functional Materials, 2017. **27**(19): p. 1603756.
27. McDonnell, S., et al., *Hole contacts on transition metal dichalcogenides: Interface chemistry and band alignments.* ACS nano, 2014. **8**(6): p. 6265-6272.
28. Grau-Crespo, R., et al., *Modelling a linker mix-and-match approach for controlling the optical excitation gaps and band alignment of zeolitic imidazolate frameworks.* Angewandte Chemie, 2016. **128**(52): p. 16246-16250.
29. Hill, H.M., et al., *Band alignment in MoS2/WS2 transition metal dichalcogenide heterostructures probed by scanning tunneling microscopy and spectroscopy.* Nano letters, 2016. **16**(8): p. 4831-4837.
30. Mak, K.F., et al., *Atomically thin MoS 2: a new direct-gap semiconductor.* Physical review letters, 2010. **105**(13): p. 136805.
31. Shen, T., A.V. Penumatcha, and J. Appenzeller, *Strain engineering for transition metal dichalcogenides based field effect transistors.* ACS nano, 2016. **10**(4): p. 4712-4718.
32. Dadgar, A., et al., *Strain engineering and Raman spectroscopy of monolayer transition metal dichalcogenides.* Chemistry of Materials, 2018. **30**(15): p. 5148-5155.
33. Liu, H., et al., *Strain engineering the structures and electronic properties of Janus monolayer transition-metal dichalcogenides.* Journal of Applied Physics, 2019. **125**(8): p. 082516.
34. Dong, R., et al., *The Intrinsic Thermodynamic Difficulty and a Step-Guided Mechanism for*

*the Epitaxial Growth of Uniform Multilayer MoS2 with Controllable Thickness.* Advanced Materials, 2022: p. 2201402.

35. Kimoto, T., *Bulk and epitaxial growth of silicon carbide.* Progress in Crystal Growth and Characterization of Materials, 2016. **62**(2): p. 329-351.
36. Nahhas, A., H.K. Kim, and J. Blachere, *Epitaxial growth of ZnO films on Si substrates using an epitaxial GaN buffer.* Applied Physics Letters, 2001. **78**(11): p. 1511-1513.
37. Liu, F., et al., *Electric field effect in two-dimensional transition metal dichalcogenides.* Advanced Functional Materials, 2017. **27**(19): p. 1602404.
38. Sharma, M., et al., *Strain and electric field induced electronic properties of two-dimensional hybrid bilayers of transition-metal dichalcogenides.* Journal of Applied Physics, 2014. **116**(6): p. 063711.
39. Murthy, A.A., et al., *Direct visualization of electric-field-induced structural dynamics in monolayer transition metal dichalcogenides.* ACS nano, 2020. **14**(2): p. 1569-1576.
40. Kresse, G. and J. Furthmüller, *Efficient iterative schemes for ab initio total-energy calculations using a plane-wave basis set.* Physical review B, 1996. **54**(16): p. 11169.
41. Blöchl, P.E., *Projector augmented-wave method.* Physical review B, 1994. **50**(24): p. 17953.
42. Perdew, J.P., K. Burke, and M. Ernzerhof, *Generalized gradient approximation made simple.* Physical review letters, 1996. **77**(18): p. 3865.
43. Krukau, A.V., et al., *Influence of the exchange screening parameter on the performance of screened hybrid functionals.* The Journal of chemical physics, 2006. **125**(22): p. 224106.
44. Kresse, G. and J. Furthmüller, *Efficiency of ab-initio total energy calculations for metals and semiconductors using a plane-wave basis set.* Computational materials science, 1996. **6**(1): p. 15-50.
45. Monkhorst, H.J. and J.D. Pack, *Special points for Brillouin-zone integrations.* Physical review B, 1976. **13**(12): p. 5188.
46. Liechtenstein, A., V.I. Anisimov, and J. Zaanen, *Density-functional theory and strong interactions: Orbital ordering in Mott-Hubbard insulators.* Physical Review B, 1995. **52**(8): p. R5467.
47. Liu, Z., B. Wang, and C. Cazorla, *Mechanical and electronic properties of CeO2 under uniaxial tensile loading: A DFT study.* Materialia, 2021. **15**: p. 101050.
48. Grimme, S., et al., *A consistent and accurate ab initio parametrization of density functional dispersion correction (DFT-D) for the 94 elements H-Pu.* The Journal of chemical physics, 2010. **132**(15): p. 154104.
49. Liu, Z., et al., *First-principles high-throughput screening of bulk piezo-photocatalytic materials for sunlight-driven hydrogen production.* Journal of Materials Chemistry A, 2022. **10**(35): p. 18132-18146.
50. Nørskov, J.K., et al., *Trends in the exchange current for hydrogen evolution.* Journal of The Electrochemical Society, 2005. **152**(3): p. J23.
51. Atkins, P., P.W. Atkins, and J. de Paula, *Atkins' physical chemistry.* 2014: Oxford university press.
52. Backes, C., et al., *Functionalization of liquid‐exfoliated two‐dimensional 2H‐MoS2.* Angewandte Chemie International Edition, 2015. **54**(9): p. 2638-2642.
53. Jaramillo, T.F., et al., *Identification of active edge sites for electrochemical H2 evolution from MoS2 nanocatalysts.* science, 2007. **317**(5834): p. 100-102.

54. Huang, Y., et al., *The reaction mechanism with free energy barriers for electrochemical dihydrogen evolution on MoS2.* Journal of the American Chemical Society, 2015. **137**(20): p. 6692-6698.
55. Tsai, C., et al., *Theoretical insights into the hydrogen evolution activity of layered transition metal dichalcogenides.* Surface Science, 2015. **640**: p. 133-140.
56. Li, M., et al., *Hydrazine Hydrate Intercalated 1T-Dominant MoS2 with Superior Ambient Stability for Highly Efficient Electrocatalytic Applications.* ACS Applied Materials & Interfaces, 2022. **14**(14): p. 16338-16347.
57. Kang, J., et al., *Band offsets and heterostructures of two-dimensional semiconductors.* Applied Physics Letters, 2013. **102**(1): p. 012111.
58. Stengel, M., N.A. Spaldin, and D. Vanderbilt, *Electric displacement as the fundamental variable in electronic-structure calculations.* Nature Physics, 2009. **5**(4): p. 304-308.
59. Cazorla, C. and M. Stengel, *Electrostatic engineering of strained ferroelectric perovskites from first principles.* Physical Review B, 2015. **92**(21): p. 214108.